\documentclass[usegraphicx,useAMS,usenatbib]{mn2e}
\newcommand{\be}{\begin{equation}}
\newcommand{\ee}{\end{equation}}
\newcommand{\bd}{\begin{displaymath}}
\newcommand{\ed}{\end{displaymath}}

\usepackage{times}

\title[Chromospheric activity in NGC~2516]
 {Chromospheric activity among fast rotating M-dwarfs in the open cluster NGC~2516}

\author[R. J. Jackson and R. D. Jeffries]
  {R. J.~Jackson and R. D.~Jeffries\\
  Astrophysics Group, Research Institute for the Environment, Physical
  Sciences and Applied Mathematics, Keele University, \\ Keele, 
      Staffordshire ST5 5BG
}
\date{MNRAS in press}

\pagerange{\pageref{firstpage}--\pageref{lastpage}} \pubyear{2009}

\def\LaTeX{L\kern-.36em\raise.3ex\hbox{a}\kern-.15em
    T\kern-.1667em\lower.7ex\hbox{E}\kern-.125emX}

\begin{document}
\label{firstpage}
\maketitle

\begin{abstract}
We report radial velocities (RVs), projected equatorial velocities
($v\sin i$) and Ca\,{\sc ii}~triplet (CaT) chromospheric activity
indices for 237 late-K to mid-M candidate members of the young open
cluster NGC~2516.  These stars have published rotation periods between
0.1 and 15\,days. Intermediate resolution spectra were obtained using
the Giraffe spectrograph at the Very Large Telescope.  Membership was
confirmed on the basis of their RVs for 210 targets. For these
stars we see a marked increase in the fraction of rapidly rotators
as we move to cooler spectral types. About 20 per cent of M0--M1
stars have $v \sin i >15$\,km\,s$^{-1}$, increasing to 90 per cent of
M4 stars. Activity indices derived from the first two lines of the CaT
(8498\AA\ and 8542\AA) show differing dependencies on rotation period
and mass for stars lying above and below the fully convective
boundary. Higher mass stars, of spectral type K3--M2.5, show
chromospheric activity which increases with decreasing Rossby number
(the ratio of period to convective turnover time), saturating for
Rossby numbers  $<0.1$.  For cooler stars, which are probably fully
convective and almost all of which have Rossby numbers $<0.1$, there is
a clear decrease in chromospheric activity as $(V-I)_0$ increases,
amounting to a fall of about a factor of 2--3 between spectral types
M2.5 and M4. This decrease in activity levels at low Rossby numbers is not seen in
X-ray observations of the coronae of cluster M-dwarfs or of active
field M-dwarfs. There is no evidence for supersaturation of
chromospheric activity for stars of any spectral type at Rossby numbers
$<0.01$. We suggest that the fall in the limiting level of
chromospheric emission beyond spectral type M3 in NGC~2516 is, like the
simultaneous increase in rotation rates in field stars, associated with a change in
the global magnetic topology as stars approach the fully  convective boundary
and not due to any decrease in dynamo-generated magnetic flux.
\end{abstract}

\begin{keywords}
 stars: rotation -- stars: magnetic activity; stars: low-mass --
 clusters and associations: NGC 2516. 
\end{keywords}

\section{Introduction}
Measurements of the masses and radii of M-dwarfs are significantly discrepant
from the predictions of evolutionary models (Ribas et
al. 2008). Initial evidence for this comes from eclipsing binaries,
where radii are 10--15 per cent higher at a given mass than predicted
(L\'opez-Morales 2007; Morales et al. 2009). It has been suggested that
the presence of dynamo-generated magnetic fields in what are relatively
fast rotating stars, can suppress convection, produce cool star spots
and hence reduce the stellar effective temperature (D'Antona, Ventura
\& Mazzitelli 2000; Mullan \& MacDonald 2001; Chabrier, Gallardo \& Baraffe
2007). Jackson, Jeffries \& Maxted (2009) measured the radii of
single, rapidly rotating M-dwarfs in the young open cluster
NGC~2516. They found that their radii, at a given luminosity, are
also larger than predicted by evolutionary models. The
discrepancy increases from a few per cent for early (M0) M-dwarfs, to
some 50 per cent for mid-M dwarfs ($\simeq $M4). The same evolutionary
models correctly predict the radii of magnetically inactive M-dwarfs,
thus implicating rotationally induced magnetic activity as the source
of the discrepancy.  Whilst this appears credible in qualitative terms,
further data are required to correlate measurements of mass and radii
with measurements of rotation, magnetic field strength and indicators
of chromospheric and coronal activity.

In low-mass F-, G- and K-type stars, the ratio of coronal X-ray to
bolometric flux, $L_{\rm x}/L_{\rm bol}$, or a variety of similarly defined
chromospheric flux indicators, are used as proxies for magnetic
activity. Magnetic flux and X-ray/chromospheric activity both appear to
depend primarily on rotation rate, but also on the convective turnover
time, as expected from simply dynamo models (e.g. Mangeney \& Praderie
1984).  Magnetic flux and magnetically induced emissions increase with
rotation speed and with decreasing Rossby number -- defined as the
ratio of rotation period to convective turnover time ($N_R =
P/\tau_c$). However, for $N_R < 0.1$, magnetic activity reaches a
saturation plateau where $L_{\rm x}/L_{bol} \simeq 10^{-3}$ (Stauffer et
al. 1994). Similar saturation plateaus are also found in chromospheric
emission lines (Soderblom et al. 1993; James \& Jeffries 1997). More
limited observational evidence shows that for extremely fast rotating
G- and K-stars with $v \sin i$ in the range 100 to 200\,km\,s$^{-1}$ (and hence
$N_R\simeq 0.01$) there is a downward trend in $L_{\rm x}/L_{bol}$ from the
saturation plateau (Pizzolato et al. 2003). This effect has been dubbed
supersaturation by Prosser et al. (1996).

Less is known about the behaviour of magnetic activity in fast rotating
M-dwarfs, but these may harbour crucial clues to the explanation of
saturation and supersaturation. They have much deeper convection zones
than hotter stars and as a result have longer convective turnover times
and hence lower Rossby numbers at the same rotation
period. Furthermore, main-sequence M-dwarfs with spectral types of M3
and cooler are probably fully convective (Siess, Dufour \& Forestini
2000), so a dynamo operating at the interface between radiative core
and convective envelope can no longer explain their magnetic activity.

Rotation and magnetic activity in low-mass M-dwarfs have been the focus
of much recent work. The key points are that in field M-dwarfs there
appears to be an abrupt change in rotational properties at spectral
type M3. Hotter than this there are few rapid rotators, but for cooler
stars there are many fast rotators and almost no slowly rotating stars
(Delfosse el al. 1998; Reiners \& Basri 2008; Jenkins et al. 2009). This has been interpreted
as a rapid lengthening of the spin-down timescale, roughly coinciding
with the transition to fully convective stars. Field M-dwarfs earlier
than M3 show clear evidence of a rotation-activity relation similar to
F-K stars for coronal indicators, including
saturation at small Rossby numbers (Pizzolato et al. 2003; Kiraga \&
Stepien 2007). For fully convective stars the data are sparse. Most
fully convective field M-dwarfs are likely to have very small Rossby
numbers. Their coronal activity appears to saturate at $L_{\rm x}/L_{\rm bol}
\simeq 10^{-3}$ out to spectral types of at least M6 (Delfosse et
al. 1998; Reiners, Basri \& Browning 2009), but peak levels of
chromospheric emission, as measured by $L_{{\rm H}\alpha}/L_{\rm bol}$,
are lower in stars cooler than M6 and may begin to decline at spectral
type M4 (Mohanty \& Basri 2003). 

Supersaturation either in coronal or chromospheric indicators is
largely uninvestigated for M-dwarfs, though tentative evidence for the
effect has been claimed at X-ray wavelengths by James et al. (2000) and
by Reiners \& Basri (2010) for the chromospheric H$\alpha$ emission of
rapidly rotating M7-M9 dwarfs.

A difficulty in most of these studies is that they (understandably)
concentrate on nearby field M-dwarfs.  However, this inevitably leads
to samples with a range of ages and metallicities, a wide spread of
rotation velocities and often no information about rotation periods. A
complementary approach is to target M-dwarfs in open clusters which are
presumably coeval and chemically homogeneous and where ages and
chemical compositions have already been determined from higher mass
stars. The disadvantage here is the distance, but this can be mitigated
using multiplexing instruments which operate over a significant area
within a cluster -- for example, simultaneous time-series montitoring
of many M-dwarfs to find rotation periods or fibre spectroscopy of many
targets in one exposure.

In this paper we report the results of intermediate resolution
spectroscopy, over the wavelength range 8060\AA~to 8600\AA, for 237
late-K to mid-M dwarfs.  These are all photometric candidate members of
the open cluster NGC~2516 with published rotation periods and an age of
$\sim 150$\,Myr (Irwin et al. 2007). The spectra were analyzed to
identify cluster members and measure radial velocities (RVs), projected
equatorial velocities ($v \sin i$) and chromospheric activity using the
CaT lines.  In section~2 we review the properties of NGC~2516. In
sections 3 and 4 we report on our target selection and the observations
and data analysis of fibre spectroscopy taken with the Very Large
Telescope (VLT). Section 5 discusses the selection of cluster members
from our data and presents their rotational properties.  In section 6
the strengths of two of the CaT lines are used to determine levels of
chromospheric activity. In section 7 the results are investigated with
respect to spectral type, period and Rossby number to look for evidence
of chromospheric saturation or supersaturation.  In section 8 the
results are compared with other observations and 
discussed in the context of current theories for the
generation of magnetic fields and the magnetic topology in stars with masses
above and below the fully convective boundary.

\section{NGC~2516}
NGC~2516 is a relatively close and well studied, young open
cluster. The first reliable estimates of key parameters were by Cox
(1955) who analyzed magnitudes and colours of 166 stars in NGC~2516 to
estimate a distance of $400 \pm 25$\,pc and reddening,
$E(B-V)=0.11$. More recent papers have described membership surveys and
characterisation of the cluster mass function (Jeffries, Thurston \&
Hambly 2001; Sung et al. 2002; Moraux, Bouvier \& Clark 2005). The age
of the cluster has been determined as $\simeq$150~Myr from the nuclear
turn off in high mass stars and the lithium depletion and X-ray
activity seen in cooler stars (Jeffries, James \& Thurston 1998; Lyra
et al. 2006). Metallicity is close to solar; being determined
spectroscopically as [Fe/H]\,$= 0.01 \pm 0.07$ and photometrically as
[M/H]=\,$-0.05 \pm 0.14$ (Terndrup et al. 2002). The same authors give
an intrinsic distance modulus of $7.93 \pm0.14$ based on main sequence
fitting and a cluster reddening of $E(B-V) = 0.12 \pm 0.02$. These
values of distance modulus and reddening are used in this paper. The
distance modulus is higher than the value of $7.68 \pm0.07$ determined
from the new Hipparcos catalogue (van Leeuwen 2009) but the difference
is not critical for this paper.

The cluster contains a large population of low mass stars (Hawley,
Tourtellot \& Reid 1999, Jeffries et al. 2001). These cool stars show
similar levels of chromospheric H$\alpha$ emission to those in the
Pleiades (Hawley et al. 1999) indicating significant magnetic
activity. Irwin et al. (2007) described a monitoring survey of
low mass stars in NGC~2516, reporting rotation periods for 362 candidate
cluster members in the mass range 0.15 to 0.7 $M_\odot$. 
A number of X-ray surveys have been carried out. The most
recent and sensitive used a deep {\it XMM-Newton} observation to probe low
mass members of the cluster down to spectral type M5 (Damiani el
al. 2003; Pillitteri et al. 2006).

\begin{table}
	\caption{Details of observing program 380.D-0479}
		\begin{tabular}{cllcc}
		Ref No. & Date \& & Field Centre & Seeing & No. of\\
		Run No. & start time & RA / Dec & Airmass & targets \\\hline
		 287508 & 27-11-07 & 118.639  & 0.79& 45 (45)\\
		 1& 04:57 & -60.804 & 1.50 & \\
		 287510 & 27-11-07 & 118.209 & 0.83 & 22 (21)\\
		 2& 05:53 & -61.107 &  1.35 & \\
		 287512 & 27-11-07 & 119.102  & 0.74 & 44 (43)\\
		 3& 06:48 & -61.108 &  1.36 & \\
		 repeat & 27-11-07 & 119.102  & 0.90 & 44 (0)\\
		 4& 07:39 & -61.108  & 1.26 & \\
		 287514 & 02-01-08 & 120.010  & 1.09 & 38 (38)\\
		 5& 06:25 & -61.265  & 1.27 & \\
		 287516 & 29-11-07 & 119.804 &  1.25 & 55 (50)\\
		 6& 06:51 & -60.934 & 1.24 &  \\
		 287518 & 02-01-08 & 120.594  & 1.16 & 41 (41)\\
		 7& 05.31 & -60.790  & 1.24 & \\
		 287520 & 30-12-07 & 120.097  & 0.90 & 39 (38)\\
		 8& 02:40 & -60.549  & 1.24 & \\
		 287522 & 30-12-07 & 119.462  & 1.21 & 32 (22)\\
		 9& 01:44 & -61.402  & 1.76 & \\ \hline
		 \multicolumn {5} {l} {The field centres are in degrees and average seeing in arcseconds.}\\
		 \multicolumn {5} {l} {Values in brackets indicate the number of new targets observed.}
		 \end{tabular}
	\label{tab:OBs}
\end{table}

\section{Spectroscopic Observations}

The spectroscopy targets were chosen from candidate NGC~2516
members reported by Irwin et al. (2007). These have rotation periods in
the range 0.1 to 15 days and we selected from those in
the range 14.5 $<$ I $<$ 18.5, corresponding to an approximate mass
range of $0.2 < M/M_ \odot < 0.7$ (Jackson et al. 2009). The targets were
observed using the European Southern Observatory (ESO) 8.2m aperture
Very Large Telescope (UT-2 Kueyen) FLAMES fibre instrument, feeding the
Giraffe and UVES spectrographs. Between 22 and 55 of our targets
were observed with Giraffe in each of eight separate fibre
configurations (see Table~1). In each configuration,
about 15 fibres were placed on ``blank'' sky positions and
another $\sim 40$ were placed on other photometric candidate members of
NGC~2516 with unknown rotation period.  The Giraffe spectrograph was
used with the HR20A grating, covering the wavelength range
8060-8600\AA\ at a resolving power of 16\,000. At least two bright,
early-type stars in the same field of view were simultaneously observed
using fibres feeding the UVES spectrograph
at a resolving power of 47\,000. 

\begin{figure}
	\centering
		\includegraphics[width = 85mm]{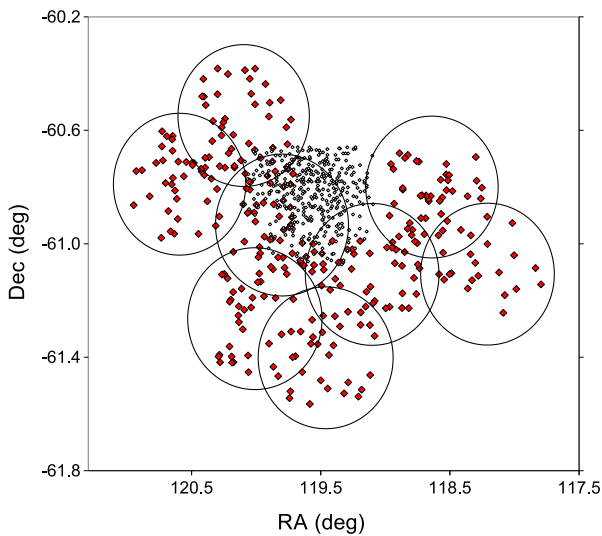}
	\caption{Coordinates of targets in the open cluster NGC~2516. The
	solid diamonds show target stars from Irwin et al. (2007) 
        with known periods and with spectra that were
	measured in 8 separate FLAMES configurations (see text). 
        The small circles show the
	coordinates of {\it XMM-Newton} X-ray sources 
        (from Table A.1 of Pillitteri et al. 2006).}
	\label{fig1}
\end{figure}

Details of the eight configurations are shown in Table~1 and their
spatial locations indicated in Fig.~1.  All targets were located within
25 arcmin of the field centres. Each configuration was observed with
two sequential 1280\,s exposures with Giraffe and three 800\,s exposures
with UVES. One configuration was observed twice during the same
night. These repeated data were useful in assessing measurement
uncertainties.  A total of 360 spectra were recorded for 294 unique
primary targets. Many targets were observed on more than one occasion,
either in the repeated observation mentioned above or through different
fibres in another, overlapping configuration (see Fig.~1).

It is worth noting in Fig.~1 that Irwin et al. (2007) concentrated
their observations in the outskirts of NGC~2516 to avoid problems
associated with bright, high-mass stars near the cluster centre. As a
result there is little overlap between our target list and the X-ray
sources found in the sensitive {\it XMM-Newton} observation of
Pillitteri et al. (2006).

\subsection{Extraction of target spectra}
Many of the target spectra were faint, requiring optimal extraction to
provide sufficient signal-to-noise ratio (SNR) for useful analysis.
There was also strong telluric absorption and sky emission lines
present. For these reasons we used our own purpose-built software for data
reduction of the Giraffe spectra. However, the high SNR spectra produced from
the software pipeline provided by ESO were found to be satisfactory for
the telluric reference stars observed with UVES.

For the Giraffe data, images of the target fields and associated flat and arc
exposures were debiased and rebinned to compensate for the initial curvature of
the spectra on the CCD image. A median of 15
bias images was used for all compensation, with overscan regions used
to correct for any time-dependent bias level. The flat frames were
the median of three tungsten-lamp flat exposures recorded each day
prior to night-time observations. A mask frame was prepared using a
3600\,s dark frame. We rejected pixels accumulating more than 5 counts
during this period, which appeared to eliminate bad pixels and also a small ``hot''
region in one corner of the CCD.

One dimensional spectra were extracted from the science frames using
the procedure described by Horne (1986). This applies a non-uniform
weight to pixels in the extraction sum, minimising statistical noise
whilst preserving photometric accuracy. Horne showed that the ideal
weighting function is the normalised spatial profile of the image at a
given wavelength. In our case the spatial profile of the image was
determined from a boxcar average of the flat field spectrum taken over
20 pixels in the wavelength direction. This averages out local
variations in the response of individual pixels whilst still allowing
the averaged profile to follow minor variations in the horizontal ($x$)
position of the spectrum centre with wavelength. Science and flatfield
spectra were extracted separately and the ratio taken to compensate for pixel-to-pixel
response variations.

Arc spectra were extracted from Thorium-Argon
lamp exposures recorded during the day prior to an observation. Gaussian
fits were used to determine the locations of 13 well separated,
unsaturated lines in the arc spectrum recorded through each fibre. Cubic
polynomial fits to these were used to bin spectra onto a
wavelength scale between 8061\AA\ and 8614\AA~in 0.05\AA~steps. 
A fine adjustment was made to the wavelength scale of each
observation through each fibre to compensate
for time-dependent drift. The adjustment was
determined by comparing the positions of emission lines in the median
sky spectra of each observation to their positions in sky spectra
averaged over all observations. The magnitude of the offset varied between
0.001\AA~and 0.008\AA~(0 to 0.3~km\,s$^{-1}$).

Finally, target spectra were sky subtracted, averaged and corrected for
telluric absorption. To allow for variations in fibre efficiency, the
proportion of the median sky spectrum subtracted from each target
spectrum was tuned to minimise the peak amplitude of the
cross-correlation function between the sky-subtracted target spectrum
and the median sky spectrum. Additionally, the sky subtracted spectrum
was masked over the width of the major sky emission lines. The two
exposures recorded within each observation block were averaged if they
differed by less than $2\sigma$ otherwise the lower of the two local
values was taken (to deal with cosmic rays). Spectra from each
configuration were corrected for telluric absorption over the
wavelength 8060\AA~to 8440\AA~using templates derived from the
co-temporal UVES spectra of bright blue stars, which were broadened to
mimic the spectral resolution of Giraffe. The
broadened spectra showed only minor effects of telluric absorption for
$\lambda > 8440$\AA, so no telluric compensation was
made in this wavelength range.

\section{Radial and Projected Equatorial Velocities}

To determine radial velocities (RV) and projected equatorial velocities
($v\sin i$), the spectra of target stars were rebinned onto a
logarithmic scale and convolved with template spectra of standard stars
over the wavelength range 8061\AA~to 8530\AA, deliberately avoiding the
chromospherically contaminated CaT lines and masking out major sky
emission features.  Templates of type K4.5V (HD~209100) and M6V
(HD~34055) from the UVES atlas (Bagnulo et al. 2003) were used,
encompassing the full range of target spectral types. These were
broadened using a Gaussian kernel to match the resolution of the
Giraffe spectra. A Gaussian profile, was fitted to the peak in the
cross-correlation function over a width of $\pm0.8 \sigma$. The offset
of this profile gave the relative RV of the target star and its width,
$\sigma$, gave a measure of rotational broadening that we used to
estimate $v \sin i$. For spectra with a SNR
$\geq 5$, there was usually a clearly defined, single peak in the
cross-correlation function that could be fitted with a Gaussian profile
to determine a unique RV and $v \sin i$. For spectra with average SNR
$< 5$, the cross-correlation peak was often sufficiently distorted by
random noise that we considered the results unreliable and these
targets were rejected from our sample at this stage.

\begin{table*}
\caption{Velocity data measured for targets in the open cluster
NGC~2516.  The identifier, co-ordinates and periods of targets are
taken from Table 1 of Irwin et al. (2007). $V$ and $I$ photometry are
also from Irwin et al., but corrected in the way described in
section~4.2.  $K$-band photometry comes from 2MASS, but is transformed
to the CIT system. Relative RV and $v \sin i$ values are given for 237
targets which have a spectral SNR$\geq 5$, 210 of which are identified as
cluster members. The right hand column indicated the run number(s) for
the observation (see Table 1), an asterisk indicates a non-member.
The full table is available on Blackwell Synergy as
Supplementary Material to the on-line version of this table.}
\begin{tabular}{lllllllrrrl} \hline 

Identifier & RA      & Dec     & Period & V     & $I_{J}$   & $K_{CIT}$ & SNR & RV     & $v \sin i$ & Run \\
    ~      & (J2000) & (J2000) & (d)    & (mag) & (mag) & (mag)   &  ~  & (km\,s$^{-1}$) & (km\,s$^{-1}$)    &Nos. \\ \hline       
N2516-1-1-1470 & 7 57 8.92 & -61 29 18.6 & ~8.803 & 17.67 & 15.49 & 13.47 & 26 & ~-0.56$\pm $0.47 &  ~$<$8.00  &  9 \\
N2516-1-1-1667 & 7 57 16.58 & -61 31 38.7 & ~1.347 & 15.74 & 14.53 & 12.72 & 44 & ~~0.16$\pm $0.46 &  21.98$\pm $1.91 &  9 \\
N2516-1-1-2490 & 7 57 47.14 & -61 30 37.5 & ~0.968 & 19.97 & 16.99 & 14.69 & 10 & ~~1.43$\pm $1.80 &  25.63$\pm $3.04 
&  9 \\
N2516-1-1-3264 & 7 58 20.21 & -61 33 53.8 & ~1.320 & 19.61 & 16.77 & 14.32 & 11 & ~~2.50$\pm $1.61 &  24.12$\pm $3.07 &  9 \\
N2516-1-1-351 & 7 56 28.22 & -61 27 46.6 & ~2.318 & 18.99 & 16.35 & 14.17 & 16 & ~~1.40$\pm $0.80 &  14.44$\pm $3.8 &  9 \\
N2516-1-1-3695 & 7 58 34.29 & -61 27 08.0 & ~9.606 & 16.95 & 14.97 & 12.97 & 33 & ~~1.77$\pm $0.41 &  ~$<$8.00  &  9  * \\
N2516-1-1-958 & 7 56 49.99 & -61 32 19.9 & ~6.291 & 18.48 & 15.99 & 13.63 & 19 & ~~1.06$\pm $0.61 &  ~$<$8.00  &  9 \\

    \end{tabular}
  \label{velocity_data}
\end{table*}

RVs were heliocentrically corrected to an arbitary zeropoint for each standard.
Values of $v\sin i$ were determined from the measured widths using
calibration curves derived by broadening standard star
spectra to simulate a series of known rotation velocities. This was
done in two stages. The standards were first broadened with a Gaussian
to match the cross-correlation function widths to the average width,
$K$, obtained for 40 slow-rotating targets (with period, $P > 5$ days),
as these were expected to have negligible broadening compared with the
spectral resolution. This gave ``zero velocity'' widths of $K=19.14\pm
0.64$~km\,s$^{-1}$ and $K=17.66 \pm 0.48$~km\,s$^{-1}$ for the K4.5 and
M6 standards respectively. 
These broadened spectra were then convolved with rotational
broadening kernels for $v \sin i$ values between 0 and 100\,km\,s$^{-1}$.  A
linear limb darkening coefficient of 0.6 was used (Claret,
Diaz-Cordoves \& Gimenez 1995), but the results are insensitive
to this parameter. Originally it was intended to interpolate between
$v \sin i$ values found from the two templates, using colour as a
proxy for spectral type. However, we found that the two $v \sin i$ values
did not differ significantly, either on
average or as a function of colour, so we took their average
to minimise any uncertainty.

\subsection{Uncertainty in RV and $v \sin i$}

Uncertainties in the RV and $v \sin i$ measurements were determined
by comparing repeated measurements made on a subset of the
targets. The functional form of the uncertainty in RV and width was 
found by analysing the uncertainty produced by convolving artificially 
broadened standards  with dummy spectra, generated by injecting random 
noise of a Gaussian distribution at  increasing levels of SNR into the 
standard spectra.This indicated an uncertainty of the form
\begin{equation}
\sigma  = \sqrt{(A+B(v \sin i)^2)^2/SNR^2 + C^2}
\label{uncertainty form}
\end{equation}
for both RV and the width $W$ of the cross-correlation function,
where $A$, $B$ and $C$ are empirically derived constants.

The constants $A$ and $B$ which characterise the effects of noise in
the measured spectra were found by comparing RVs and widths measured
for 60 targets (44 with measured periods) in the repeated runs with 
the same configuration (see Table 1). The constant $C$ which represents 
additional uncertainties due to
changes in fibre allocation and night-to-night calibration variations
was estimated by comparing results from spectra  for 16 targets recorded on
different days with different fibre allocations. The empirically
derived uncertainties in RV and cross-correaltion width
were
\begin{equation}
\sigma _{RV} = \sqrt{(9.2+0.013(v \sin i)^2)^2/SNR^2 + 0.31^2} \ \ {\rm km\,s}^{-1}
\label{uncertainty RV}
\end{equation}
\begin{equation}
\sigma _{W}  = \sqrt{(8.8+0.005(v \sin i)^2)^2/SNR^2 + 0.40^2} \ \ {\rm km\,s}^{-1}
\label{uncertainty width}
\end{equation}

The uncertainty in $v\sin i$ was then determined from its relationship
with $W$ and the ``zero velocity width'' $K$ for a particular
standard.
\begin{equation}
v \sin i  = \alpha (W-K)^{1/2}
\label{vsini_form}
\end{equation}
where $\alpha$ is a scaling constant. Hence the uncertainty in $v\sin i$ is
\begin{equation}
\sigma_{v\sin i} = v\sin i \sqrt{\sigma_W^2 + \sigma_K^2} /[2(W-K)]\ \ {\rm km\,s}^{-1}
\end{equation}

\begin{figure}
	\centering
		\includegraphics[width = 85mm]{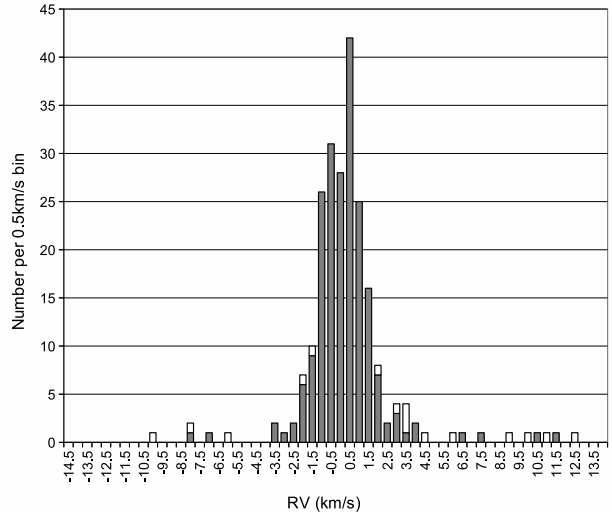}
	\caption{Histogram showing the number density of targets in
	open cluster NGC~2516 as a function of radial velocity
	relative to the mean RV of the cluster. The solid bar shows the
	count of cluster members in 0.5km\,s$^{-1}$ bins. The open bar shows
	the number density of targets classified as non-members (see text).}
	\label{fig2}
\end{figure}

\begin{figure}
	\centering
		\includegraphics[width = 85mm]{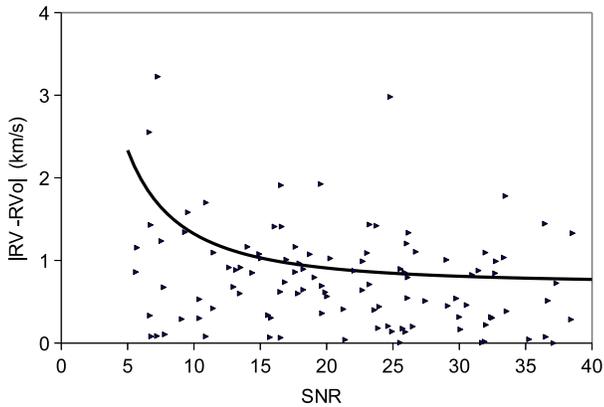}
	\caption{Variation of the radial velocity of slow-rotating members of open
	cluster NGC~2516 relative to the mean velocity of the cluster
	with signal to noise ratio of the target spectra. The solid
	line shows a best fit to the standard deviation of the RVs from
	the cluster mean as a function of SNR (see equation 7).}
	\label{fig3}
\end{figure}

\subsection{Tabulated results}

Spectra with SNR$\geq$5 were obtained for 239 stars with known
periods. One of these (N2516-3-8-1301) showed a clear double peak in the cross
correlation and is almost certainly a double-lined spectroscopic
binary. The other (N2516-1-2-369) shows a more noisy cross-correlation and is
a possible binary system.  The measured RV and $v\sin i$ for the
remaining 237 targets are given in Table 1. $V$ and $I$ photometry was
initially taken from Irwin et al. (2007), but we corrected their
photometric values to put them onto the better calibrated photometric
scale of Jeffries et al. (2001), using a set of stars common to both
papers.  The corrections added to the Irwin et al. values were $\Delta I =
0.080-0.0076\,I$ and $\Delta (V-I) = 0.300-0.153\,(V-I)$.  $K$
magnitudes are from the Two-micron All-Sky Survey (2MASS) catalogue
(Cutri et al. 2003), transformed to the CIT
system using $K_{CIT} = K_{2MASS} + 0.024$ (Carpenter 2001).

The RVs quoted in Table~1 are given relative to the average cluster RV,
estimated using the RVs of slowly rotating cluster members ($v \sin i <
20$\,km\,s$^{-1}$, see section~5.1). 
A minimum resolvable value of of 8 km\,s$^{-1}$ was taken
for $v\sin i$. At this level of broadening the increase in measured
width is $\simeq 1.3$ times the standard deviation of the ``zero
rotation'' width; hence there is a 90 per cent probability that $v\sin i$
is truly non-zero. For targets with repeated measurements the Table
shows the weighted mean of the RV and $v\sin i$ with
appropriate errors. None of the targets with repeated RV measurements
showed significant evidence for binarity.  The right hand column of
Table~1 indicates the run number(s) for the observation. The 27 targets
with periods that were identified as non-members are flagged with an
asterisk.

\section{Results}
\subsection{Membership}

In this section we aim to establish a secure list of cluster members
based on photometric colour and RV. Chromospheric activity and rotation
are not used as indicators of membership since we wish to study
the distribution of these parameters for cluster members. All target
stars were identified as potential cluster members by Irwin et
al. (2007) based on their position in the $V$/$V-I$ colour magnitude
diagram. We made a further selection based on RV relative
to the mean RV of cluster members.

Figure 2 shows a histogram of targets as a
function of relative RV in 0.5\,km\,s$^{-1}$ bins. These are tightly bunched,
within a few km\,s$^{-1}$, suggesting that the majority of the 237 targets are
indeed cluster members. For the purposes of this paper cluster members were
defined conservatively as those stars with a measured RV less than $2\sigma_e$
from the mean RV, where $\sigma_e$ is the effective velocity dispersion
due to the combined effects of RV uncertainty and the
true velocity dispersion of the cluster.  Assuming the same functional
form as equation 1 then the total dispersion will vary as
\begin{equation}
\sigma_e = \sqrt{(A+B(v \sin i)^2)/SNR^2 + C^2 + \sigma_c^2}\, ,
\end{equation}
where $\sigma_c$ is the intrinsic velocity dispersion of cluster stars.
Figure~3 shows a plot of the modulus of the relative RV against SNR for
the slower rotators ($v \sin i <20$\,km\,s$^{-1}$). Taking the mean RV as the
average value clipped at $\pm $5~km\,s$^{-1}$ then a least squares fit to the
relative RV of the slower rotators gave
\begin{equation}
\sigma _e = \sqrt{(11.1 \pm 1.4)^2/SNR^2 + (0.72\pm 0.17)^2}  
\label{dispersion_form}
\end{equation}

The constant term 0.72 $\pm$ 0.17~km\,s$^{-1}$ places an upper bound on the
intrinsic velocity dispersion for M-dwarfs in the cluster. The constant
$C$ was estimated by comparing RVs for repeated measurements made on
different days with different fibre configurations. This gave a value
of $C\simeq 0.31$~km\,s$^{-1}$ (see equation 2) and hence our best estimate for
the true velocity dispersion of the cluster is 0.66$\pm$0.17~km\,s$^{-1}$.

Using equation 7 gave 210 probable cluster members with a SNR$\geq$5
and a relative RV less than $2\sigma_e$ from the mean (see Fig.~2). A
few fast rotating stars with a correspondingly large RV uncertainty
were identified as members even though their RVs are some distance from
the mean in absolute terms. There is no reason to doubt their
membership since fast rotators are rare amongst late-K to mid-M dwarf
field stars (Delfosse et al. 1998). Also shown in Fig.~2 is the number
of stars classified as non-members. Nine of the non-members lie between
$2\sigma_e$ and $3\sigma_e$ of the mean so could yet be cluster members
(possibly SB1 binary systems), as for a Gaussian distribution we expect
$\sim 10$ further members to lie beyond $2 \sigma_e$. Others may be be
background or foreground stars. To estimate the maximum number of
background stars falsely classified as members, the average number of
targets in $\pm 2\sigma_e$ bins centered at $\pm 10$~km\,s$^{-1}$ from
the cluster mean was counted. On average these bins contained 3 non-members
with periods. Assuming that the distribution of non members is
uniform with RV this indicates that $\leq 3$ of the 210
stars identified as members are likely to be non-members.

\begin{figure*}
	\centering
	\begin{minipage}[t]{0.95\textwidth}
	\includegraphics[width = 160mm]{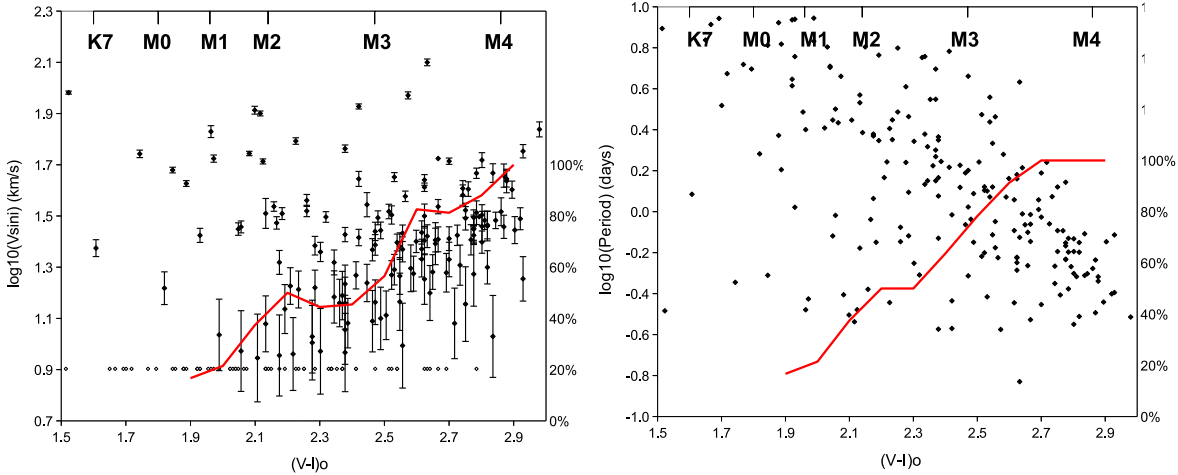}
	\end{minipage}
	\caption{The distribution of projected equatorial velocities ($v
	\sin i$) and rotation period with colour for open cluster NGC~2516. Solid
	diamonds show the measured values for objects considered to be
	cluster members (see section 5.1). Open
	diamonds show $v\sin i$ upper limits assuming a minimum resolution of
	8~km\,s$^{-1}$. The solid lines indicate (on the right hand $y$-axis
	scale) the fraction of $v \sin i$ values exceeding
	15~km\,s$^{-1}$ or the fraction of rotation periods smaller than 2~days
	respectively.}
	\label{fig4}
\end{figure*}

\subsection{Rotation rates for cluster members}
Figure 4 shows the distribution of of $v \sin i$ and period 
as a function of $(V-I)_0$ colour
for cluster members with measured periods. Colour is corrected for
reddening assuming a uniform reddening of $E(B-V)=0.12$ (Terndrup et
al. 2002) and ratios of selective to total extinction $A_V/E(B-V) =
3.09$  (Reike \& Lebofsky 1985) and $A_I/E(V-I)=2.35$. 
Spectral types are shown using the calibration from 
Kenyon \& Hartmann (1995).

For the late-K and early-M stars, the distribution of $v \sin i$ and
period are similar to those seen in other clusters at a similar age
(e.g. the Pleiades, see Queloz et al. 1998, Terndrup et
al. 2000). There is a wide range of rotation rates and periods, with
significant populations of slowly rotating stars (periods of a few days
and $v \sin i$ unresolved) and a tail of fast rotators with $v \sin i$
between 50 and 100\,km\,s$^{-1}$.

Rotation velocities appear to increase (a decrease in
period) as we move to later M-type stars. Of course, the presence of
lower limits to $v \sin i$ means an average cannot be calculated
directly. Instead, Fig.~4 shows (on the right-hand $y$-axes) the
fraction of rapid rotators with $v \sin i>15$\,km\,s$^{-1}$ or period $<2$
days, as a function of colour in 0.1~mag bins. These two criteria are
approximately equivalent for stars with a radius of 0.6\,$R_{\odot}$.
The plots show that the fraction of rapid rotators is a sharply
increasing function of colour (or decreasing mass). About 90 per cent
of stars of spectral type $\geq$ M4 are rapid rotators, compared to
about 50 per cent at M3 and only $\sim 20$ per cent at M0-M1. 

This spectral type dependence appears to set in during the first few
Myr of stellar evolution and develop slowly over time (Irwin et
al. 2007). Samples of field stars, presumably with ages measured in
Gyr, have almost no fast rotators earlier than type M3, but a sharp
increase in the fraction of rapid rotators among cooler stars (Delfosse
et a. 1998; Jenkins et al. 2009; Browning et al. 2010). West et
al. (2008) have calibrated age-magnetic activity relationships for
field M-dwarfs, suggesting that while M0 dwarfs have an
``activity lifetime'' of 0.8\,Gyr, this increases to 4.5\,Gyr for M4 dwarfs.
Magnetic activity is closely related to rotation rate, so it is not
surprising that the $\simeq 150$\,Myr old M0--M2 stars in NGC~2516
rotate much faster on average than field stars (where the majority of
field stars would have spun down), but have similar rotation rates to
the cooler field stars, which have not had time to spin down.

The usual interpretation of these phenomena (e.g. Delfosse et al. (1998; Jenkins et al. 2009
) is that spin-down timescales become much
longer for cooler stars because the change from stars with
radiative cores to fully convective stars changes the
magnetic topology and makes angular momentum loss less efficient.  As the
vast majority of targets with rotation periods have been confirmed here
as cluster members, then we refer the reader to Irwin et al. (2007),
where the rotation period distribution is modelled in some detail in
terms of angular momentum loss from a magnetized stellar wind.

\section{Calcium triplet lines}

The wavelength range of our spectra (8061--8614\AA) includes two of the
CaT lines at rest wavelengths of 8498\AA~and 8542\AA.  The CaT shares
an upper level with the better known Ca~{\sc ii}~H~and~K lines, which
are more often used as chromospheric activity indicators. The CaT lines
are also known to be effective indicators of chromospheric
activity, with stars of similar luminosity and metallicity having
different CaT line depths, owing to varying levels chromospheric
emission filling the underlying absorption lines (Mallik 1994,
1997). Bus\`a et al. (2007) show that the chromospheric component of
the CaT lines is well correlated with the chromospheric Ca~{\sc
ii}~H~and~K flux.

The method used here to measure the strength of the CaT chromospheric
emission follows that described by Marsden, Carter \& Donati (2009). We estimate
the chromospheric component of the CaT flux by subtracting the
photospheric contribution from a magnetically inactive star of similar
spectral type.

\subsection{Reference spectra}
Marsden et al. (2009) used this technique to measure CaT emission in
F,G and K stars.  The difficulty in using it for M dwarfs is finding
suitable reference spectra at these spectral types with known low rotation and
chromospheric activity.  Several sources were investigated but none
alone could provide a well defined set of standards spanning the
range K3 to M5. For this reason results from several sources
were combined to generate a semi-empirical reference spectra by scaling
the measured high SNR spectrum of a magnetically inactive K3 standard by a
colour-dependent scaling factor determined from lower SNR spectra
of late-K and M dwarf stars.

\begin{figure}
	\centering
		\includegraphics[width = 85mm]{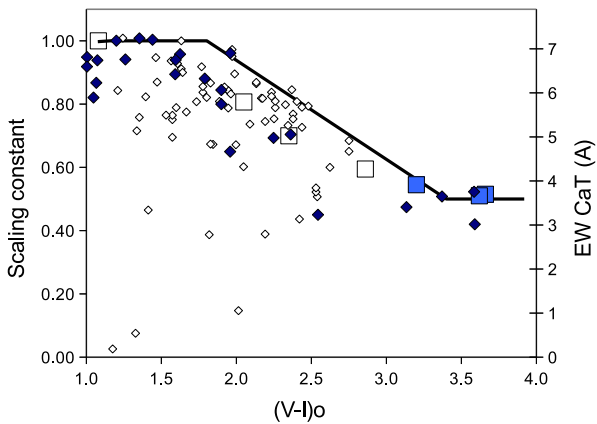}
			\caption{Variation of the scaling factor used
			to define the semi-empirical reference spectra
			as a function of $(V-I)_0$. The solid line
			defines the scaling factor as the upper limit
			of the relative depths of the first two calcium
			triplet lines measured in spectra of
			low-activity 
			stars from various sources, compared
			with the K3 reference star GJ\,105a. Open squares
			were found from spectra in Montes \& Mart\'in
			(1988), blue shaded squares from spectra of Jenkins et al. 
			(2009) and small open diamonds  from
			slowly rotating
			non-members of NGC~2516 (see text).  These EWs have been
			normalised to match up with the low activity
			stars at the blue end of the data. The solid
			diamonds show the original CaT EWs from Cenarro et
			al. (2001) and are refered to the right hand y-axis.}
	\label{fig5}
\end{figure}

The first source of reference spectra was the library of high
resolution spectra of Montes and Mart\'in (1988). This yielded
high SNR spectra for a K3 star (GJ105a), an M1.5 star (GJ15a), an M2.5
star (GJ623ab) and an M4 star (GJ748). Unfortunately only the first of
these stars, GJ105a, has a known long rotation period of 48 days, and
low chromospheric activity indicated by its Ca~{\sc ii} H and K lines
(Baliunas et al. 1995). For this reason the spectra of this K3
star, braodened to match the resolution of the target spectra,
 was chosen as the baseline for 
semi-empirical reference spectra. This broadened spectrum was then scaled
using a factor equal to the ratio of the average depths of the two CaT lines
seen in the spectra of stars of later spectral types to those seen in
the K3 standard. The scaling factor was determined as a function of $(V-I)_0$.

Figure~5 shows the scaling factor for the Montes \& Mart\'in (1988)
spectra (as large open squares). The relative depths of CaT lines for
these M-dwarf spectra are between 20 and 40 per cent lower than the K3
standard. Unfortunately these spectra cannot be directly used as
standards since their degree of activity is not known. Instead we
combined these with other spectra to try and delineate the upper
boundary of the scaling factor (as a function of colour), which should be
defined by the least chromospherically active stars.

\begin{figure}
	\centering
		\includegraphics[width = 80mm]{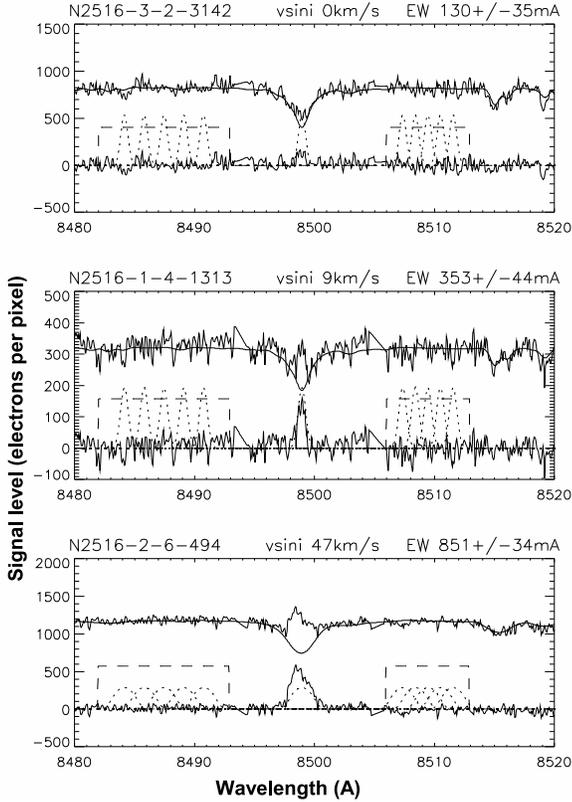}
	\caption{The process used to measure the EW of
	the CaT 8498\AA\ line shown for three representative
	spectra. The upper traces in each panel show the target spectrum
	local to the emission line together with  the (scaled) reference 
	spectrum 	(the smooth line). The lower trace in
	each panel shows the difference spectrum. Dashed lines indicate
	the sections of spectra used to normalise the target and
	reference spectra. The central dotted curve indicates the
	Gaussian extraction profile used to estimate the residual
	chromospheric EW from the difference spectra
	(equation~10). The dotted Gaussian profiles shown either side
	were used determine the uncertainty in EW (see
	text).}
\end{figure}

A second source of low activity stars are our
spectra of targets identified as non-members and for which no periods
were reported by Irwin et al. (2007). We additionally filtered these
targets on the basis that they had $v\sin i< 8$ km\,s$^{-1}$, a SNR
$>20$ and that their sodium lines at 8183\AA~and 8193\AA~had a total
equivalent width greater than 1\AA~indicating that they are dwarf stars
rather than background giants (Schiavon et al. 1997). There is a good
chance that this subsample contains a number of older, less active
stars. Figure 5 shows the relative depths of the CaT lines of these
stars as open diamonds.

To extend the range of spectral types beyond M4, additional results were 
added for three stars, (LHS\,3075, LHS\,1950 and LHS\,302) with
known low chromospheric activity ($H_{\alpha}$ not in emission) and low
measured $v\sin i$ ($<2.5$ km\,s$^{-1}$) (Jenkins et al. 2009). The
spectra for these stars (supplied by Jenkins, private communication)
were too noisy to be used directly as standards, however (after broadening 
to match the resolution of the target spectra) they could be
used to estimate the relative depth of the CaT lines compared to the
K3 standard. These results, shown as filled squares in Fig.~5 confirm
that the relative depth of the first two CaT lines falls by
$\simeq 50$ per cent between spectral types  K3 and M5, presumably
owing to depression of the local continuum by increasing molecular opacity.

Finally, we investigated the library of low resolution spectra assembled
by Cenarro et al. (2001). The resolution of these spectra (1.5A) is too low
for them to be used as standards however we measured the variation in
the total equivalent width (EW) of the CaT lines as a function of spectral
type. This is not directly comparable with the relative depth of the
CaT lines (unless the line profile is independent of colour) but should
be broadly similar. Figure 5 shows as solid diamonds the
variation of the Paschen-corrected EW of the CaT triplet taken from
Table 6 of Cenarro et al. 2001, where the V-I colours are derived from
the reported effective temperatures using the colour-temperature
relation of Kenyon \& Hartmann (1995). The EWs (shown in \AA\ on the
RH axis) are scaled to match the K3 standard at the blue end of the
plot. The results show a similar upper bound to the previous results,
indicating that the variation in scaling constant with colour is
closely related to the changes in the underlying EW of the CaT
absorption line with spectral type.

The results in Fig.~5 were used to define a semi-empirical reference
spectrum, $R(\lambda)$, as a function of colour, which represents 
chromospherically inactive stars.
Using the normalised spectrum of the K3 reference star
local to the first two CaT lines $R_{K3}(\lambda)$ and a
scaling constant, $S_c$, we define
\begin{equation}
	R(\lambda) = 1- S_c [1-R_{K3}(\lambda)]\, .
\end{equation}
We modelled $S_c$ using a simple by-eye three-part linear function to
represent the upper boundary of the points in Fig.~5, which we assume
represents the CaT line depths for chromospherically inactive stars,
where $S_c = 1$ for $(V-I)_0<1.8$ ($\simeq$ M0) and decreases linearly
to $S_c = 0.5$ for $(V-I)_0>3.4$ ($\simeq$ M5).

\subsection{Equivalent widths of the CaT emission lines}
The EWs of the first two CaT emission lines were measured
by comparing a section of the the target spectrum local to the
CaT line centre, $S(\lambda)$, with the reference spectrum, $R(\lambda
)$ (see Fig. 6). The reference spectrum was aligned in wavelength and
convolved with a broadening kernel, according to the RV and
$v\sin i$ of the target. The target and reference spectra were then
normalised to their average levels either side of the CaT line centres
over the wavelength ranges indicated in Fig. 6. The reference spectrum
was subtracted from the target spectrum to produce a difference
spectrum
\begin{equation}
	\Delta(\lambda) = S(\lambda) - R(\lambda)\, .
\end{equation}
The EW of the CaT features in $\Delta(\lambda
)$ gave a measure of chromospheric activity.

In practice the difference spectra can be noisy and the width of
the residual chromospheric component depends on the $v \sin
i$ of the target. Integrating under $\Delta(\lambda )$ would give
different estimates of EW depending on the
integration limits.  This problem was circumvented using an
``optimal extraction'' technique -- i.e. multiplying the difference
spectrum by a Gaussian profile of unit area, which represents the
expected profile of the difference spectrum produced by a chromospheric emission
line. The width of the Gaussian profile for slowly rotating stars was
found by fitting a Gaussian to the difference spectra of
25 long-period stars which showed significant CaT emission lines. This
gave a width $\sigma_0 = 0.30\pm 0.02$\AA. A Gaussian profile,
re-centred according to the target RV and broadened beyond $\sigma_0$
according to $v\sin i$, gave a function
$P(\lambda)$, which was used to determine the EW as:
\begin{equation}
	EW = \int \Delta(\lambda)P(\lambda)d\lambda \ / \int
	P(\lambda)^2 d\lambda\, .
\end{equation}
Examples of $P(\lambda)$ are shown in Fig.~6 as dotted lines centred at
the expected emission line wavelength. The uncertainty in the
EWs due to noise in the spectrum was estimated as the rms
value of the EWs measured using the same
$P(\lambda)$ centred at five wavelengths either side of the emission line.

\begin{table*}
\caption{Rotation periods, convective turnover times, Rossby numbers,
EWs and chromospheric activity indices for the first two
CaT lines (8498\AA\ and 8542\AA) in candidate members of NGC~2516. The quoted uncertainies
(see text) do not include possible systematic errors.
Identifier and period are from Irwin et al. (2007). The turnover
times were estimated from the empirical relation of Pizzolato et
al. (2003). Rossby number was calculated using periods from Irwin et
al. The right hand column indicates the run
number(s) for the observation (see Table 1) with a asterisks
marking stars falling outside our membership criteria.  The
full table is available on Blackwell Synergy as Supplementary Material
to the on-line version of this table.}

\begin{tabular}[t]{llccclcccl} \hline 
Identifier   & $log_{10}$ & Period & Turnover & $log_{10}$& EW$_{8488}$& EW$_{8542}$ & log$_{10}$ & log$_{10}$&Run \\
  & $L/L_{\odot}$ &  (days) & time  & Rossby  & (\AA)  &   (\AA) & R'$_{Ca8488}$ & R'$_{Ca8542}$& Nos.\\ 
           &    &    & (days)  & No.   &          &           &             &            \\ \hline

N2516-1-1-1470 & -1.23 & 8.803 & ~52 & -0.77 & 0.45$\pm$0.02 & 0.50$\pm$0.03 & -4.47$\pm$.02 & -4.42$\pm$.02 & 9\\
N2516-1-1-1667 & -0.88 & 1.347 & ~35 & -1.41 & 0.42$\pm$0.01 & 0.59$\pm$0.01 & -4.49$\pm$.01 & -4.34$\pm$.01 & 9\\
N2516-1-1-2490 & -1.77 & 0.968 & ~96 & -2.0 & 0.29$\pm$0.07 & 0.28$\pm$0.05 & -4.70$\pm$.05 & -4.71$\pm$.06 & 9\\
N2516-1-1-3264 & -01.7 & 1.320 & ~89 & -1.83 & 0.63$\pm$0.04 & 0.50$\pm$0.04 & -4.35$\pm$.05 & -4.45$\pm$.05 & 9\\
N2516-1-1-351 & -1.55 & 2.318 & ~75 & -1.51 & 0.49$\pm$0.04 & 0.46$\pm$0.05 & -4.45$\pm$.03 & -4.48$\pm$.03 & 9\\
N2516-1-1-3695 & -1.04 & 9.606 & ~42 & -0.64 & 0.31$\pm$0.01 & 0.37$\pm$0.02 & -4.62$\pm$.01 & -4.55$\pm$.02 & 9  *\\
N2516-1-1-958 & -1.41 & 6.291 & ~64 & -1.01 & 0.41$\pm$0.02 & 0.47$\pm$0.01 & -4.52$\pm$.03 & -4.46$\pm$.03 & 9\\
   \end{tabular}
  \label{chromospheric_data}
\end{table*}

\subsection{Chromospheric activity indices}

We define the activity index $R^{'}_{Ca}$ as the fraction of a star's
bolometric luminosity emitted from the chromosphere in a CaT line.
The conversion between the EW of the chromospheric
component of the CaT line and the flux (in erg\,cm$^{-2}$\,s$^{-1}$) 
was established by measuring the continuum flux densities in our
defined continuum windows (see Fig.~6) in K- and M-dwarfs from the
standard spectral library of Pickles (1998). Using the $V-I$ colours
tabulated by Pickles for these stars and their $V$-band fluxes we
fitted the following relationship 
\begin{equation}
\log f_{Ca} = \log EW_{Ca} - 0.4 I_0  +0.032(V-I)_0 -9.00\, ,
\end{equation}
where $I_0$ is the intrinsic Cousins $I$-band magnitude and EW$_{Ca}$ is
the EW of either of the CaT lines (the difference in
continuum levels is 0.01\,dex or less). The bolometric flux is given by 
\begin{equation}
	\log f_{\rm bol} = -0.4V - 0.4BC - 4.605
\end{equation}
where $BC$ is the $V$-band bolometric correction (Allen 1973). 
The difference in these
expressions defines the activity index
\begin{equation}
	\log(R^{'}_{Ca}) = \log EW_{Ca} + 0.432(V-I)_{0} + 0.4BC - 4.395
\end{equation}
The bolometric corrections were interpolated from the the dereddened
colours using Table~A5 of Kenyon \& Hartmann (1995). Measurement
uncertainties in the activity indices were estimated by comparing
repeated observations of the same target. Comparisons for 78 targets
showed an uncertainty of $0.53/SNR$ in $\log(R^{'}_{Ca})$ calculated for
individual lines (listed in Table~3). This reduced to $0.46/SNR$ when
the value of $\log(R^{'}_{Ca})$ was determined from the mean of the
EWs of the two CaT lines.  Additional random
uncertainties arise from the photometry and consequent
bolometric correction, but these act in opposite directions and almost
cancel (e.g. an increase in $V-I$ from 2.5 to 2.6, would decrease the
assumed $BC$ by about 0.13\,mag).

\begin{figure}
	\centering
		\includegraphics[width = 75mm]{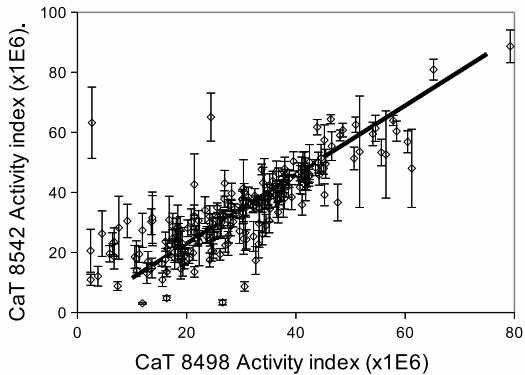}
\caption{A comparison of the activity indices measures for the first
(8498\AA) and second (8542\AA) CaT lines. Error bars show the
expected variation between repeat measurements on the same target
estimated from the spectrum SNR.}
\end{figure}

\begin{figure}
	\centering
		\includegraphics[width = 75mm]{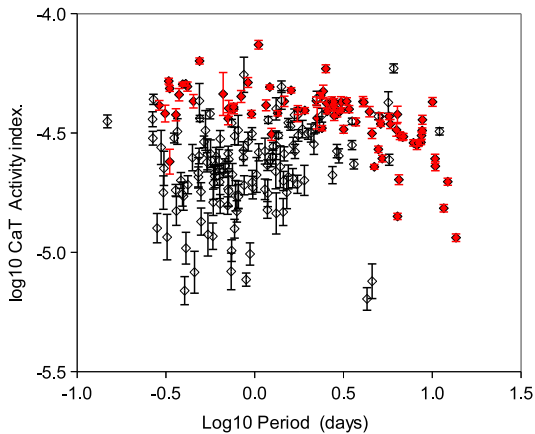}
\caption{The mean activity index of two lines of the CaT (8498\AA\ and 8542\AA) versus rotation period. 
Filled diamonds show results for targets of
spectral type K3 to M2.5; open diamonds are for spectral types M2.5 to M5.}
\label{fig9}
\end{figure}

Systematic uncertainties could also be present.  The basal photospheric
flux level in the CaT lines has been estimated from a known
magnetically inactive star in the case of the K-stars, but we have been
forced to estimate a basal photospheric flux in M-stars by looking for
the deepest photospheric CaT lines in objects with poorly constrained
activity levels. We believe this procedure is reasonably robust and
demonstrates a trend with colour that agrees with previous work on
field stars (see Fig.~5). In addition the spectra 
of a number of slowly rotating mid-M dwarfs with H$\alpha$
in absorption (and therefore presumably inactive) also lie at this
basal level. Nevertheless it is possible that all of the stars
considered in Fig.~5 have some residual level of chromospheric activity
resulting in a misleadingly weak photospheric basal level and hence
underestimated chromospheric activity at later spectral types. However,
even if $S_c$ were flat as a function of colour, this would only
increase $R'_{Ca}$ by about 0.2\,dex in the coolest M4 stars -- which
assumes an importance in the next section.

Figure 7 compares activity indices from the two CaT lines.  The results
show the expected linear correlation for activity indices above
$\approx 20 \times 10^{-6}$. The 8542\AA\ line is the
stronger by about 14 percent. The emission EWs and
chromospheric activity indices for the 8498\AA\ and 8542\AA\ CaT lines
are listed separately in Table 3. The uncertainties in the EWs
are those estimated in the extraction process.  Uncertainties in
$\log(R^{'}_{Ca})$ are as defined above. In what follows, the plotted
values of $\log(R^{'}_{Ca})$ are calculated from the mean of the two CaT lines.

\section{Magnetic activity versus rotation and spectral type}

\subsection{Chromospheric activity}

Figure 8 shows the mean chromospheric activity index plotted against
 rotation period. This plot shows considerable scatter, much more than
 expected from the measurement uncertainties, suggesting either that
 the chromospheric activity is very variable in time or that rotation
 period is not the sole parameter controlling chromospheric
 activity. If we consider results by spectral type then the earlier
 K3--M2.5 stars ($(V-I)_0<2.3$) in our sample show an initial increase in
 activity index with reducing period, which levels off for periods below
 about 3 days. Almost all stars of spectral type cooler than M2.5 have
 periods shorter than 3 days and show roughly the same constant
 activity index, but with a rms scatter of 0.2\,dex. The level of
 chromospheric activity appears to have a spectral type dependence,
 because the activity indices of the short-period K3--M2.5 stars are
 systematically higher on average than those of the $>$M2.5 sample by about
 0.3\,dex. As we discuss below, the choice of spectral type M2.5 as a
 division point is driven by the likely location of the point at which
 M-dwarfs become fully convective in NGC~2516.

The conflation of rotation and spectral type in determining activity
levels has been widely studied in F--K stars and unification
has been achieved by combining these parameters to study magnetic activity as a
function of Rossby number: $N_R = P/\tau_c$ -- the ratio of period to
convective turnover time, where the denominator is spectral-type
dependent.

The use of Rossby number raises a problem when dealing with
M-dwarfs. The widely used semi-empirical formula of Noyes et al. (1984)
predicts $\log \tau_c$ as a function of
$B-V$. This relationship is poorly defined for $B-V>1$ and has
no constraining data in the M-dwarf regime. Theoretically, little work
has been done on turnover times at very low masses.  The models of
Gilliland (1986) show that $\tau_c$ increases with decreasing mass,
from about 12 days at $1\,M_{\odot}$ to 70 days at $0.5\,M_{\odot}$.
Similar calculations, with similar results (except for arbitrary
scaling factors) have been presented more recently by Kim \& Demarque
(1996) and Ventura et al. (1998). The latter also attempted to extend
the calculation into the fully convective region, predicting that the
convective turnover time would continue to increase.

An alternative approach has been to empirically determine $\tau_c$ by
demanding that activity indicators (chromospheric or coronal) satisfy a
single scaling law with Rossby number, irrespective of stellar mass
(e.g. Noyes et al. 1984). The most recent work has focused on coronal
X-ray emission using $L_{_{\rm x}}/L_{\rm bol}$ as an activity
indicator. Using a mixture of slow- and fast-rotating stars, Pizzolato
et al. (2003) showed that $\tau_c$ needs to increase rapidly with
decreasing mass in order to simultaneously explain the behaviour of
$L_{x}/L_{\rm bol}$ in G-, K- and M-dwarfs, and they find $\tau_c >
100$~days for $M<0.5\,M_{\odot}$. Similar work by Kiraga \& Stepien
(2007) concentrated on slowly rotating M-dwarfs, finding that $\tau_c$
increases from 30~days at $M\simeq 0.6\,M_{\odot}$ to $\sim 100$~days
at $M \simeq 0.2\,M_{\odot}$. An interesting insight into this
behaviour was provided by Pizzolato et al. (2003), who noted that the
mass dependence of the turnover time is closely reproduced by
assuming it is proportional to $L_{\rm bol}^{-1/2}$.  In what follows
we follow Pizzolato et al. and adopt a Rossby number calculated from
the following equation
\begin{equation}
\log N_R = \log P - 1.1 + 0.5 \log \frac{L_{\rm bol}}{L_{\odot}}\, ,
\end{equation}
where $\tau_c$ has been anchored such that $\log \tau_c = 1.1$
for a solar-type star and $L_{\rm bol}$ is calculated from the
bolometric fluxes previously described and an assumed cluster distance
modulus of 7.93. 
The adopted $\tau_c$ values for our sample are listed
in Table~3 along with the derived Rossby numbers.

\begin{figure*}
\centering
\begin{minipage}[t]{0.9\textwidth}
\includegraphics [width = 150mm]{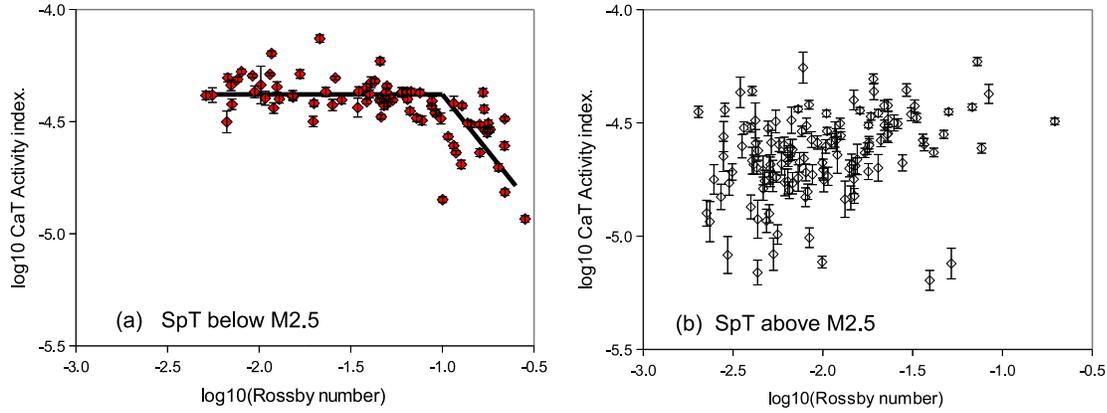}
\end{minipage}
\caption{Variation of mean chromospheric activity index for the first two lines of 
the CaT with Rossby   number (a) members of NGC~2516 with spectral types earlier than M2.5
  ($(V-I)<2.3$) and (b) for cluster members later than spectral type M2.5.}
\label{fig10}
\end{figure*}

\begin{figure*}
\centering
\begin{minipage}[t]{0.9\textwidth}
\includegraphics [width = 150mm]{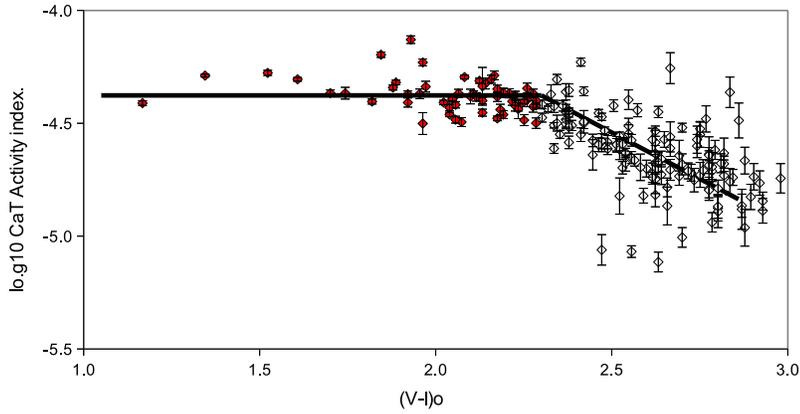}
\end{minipage}
\caption{The mean chromospheric activity index for the first two  lines of the CaT
plotted against intrinsic colour for members of NGC~2516 that have $N_R<0.1$. }
\label{fig11}
\end{figure*}

Figure~9 separately shows the behaviour of our CaT 
activity index with Rossby number for stars earlier and later than
M2.5. Unsurprisingly, the features we drew attention to in Fig.~8 are
still present, because the range in $\tau_c$ 
is only 35 to 155~days, which is considerably smaller than the factor
of 100 range in rotation periods. The stars of earlier spectral type
show the well-documented rise in activity as $N_R$ decreases
(e.g. Pizzolato et al. 2003; Kiraga \& Stepien 2007), but for $\log
N_R<-1$ the increase flattens off with, if anything, marginal evidence
for a small increase in average activity levels at $\log N_R \simeq
-2$. There is certainly no evidence for ``supersaturation'' of
chromospheric activity levels in a way analogous to that claimed for
coronal activity at $\log N_R < -1.7$ (Randich et al. 1996; Marsden et
al. 2009). 

At first glance, the behaviour of the cooler ($>$M2.5) stars in Fig.~9b
is similar. Almost all of these stars have $\log N_R <-1$ and would be
considered ``saturated'' at these Rossby numbers although it should
be noted that we have not observed any cool stars with high enough Rossby 
numbers to define an unsaturated regime. The chromospheric activity indices are
reasonably well described with a constant level of
$\log(R^{'}_{Ca})=-4.64$, albeit with a considerable and significant
scatter. A marginally better fit is achieved with an activity index
that {\it increases} with increasing Rossby number.
However, the average activity level in these stars is significantly lower, by about
0.3~dex, than that seen in the K3--M2.5 subsample with similar
$N_R$. That is, although the cooler stars appear to show ''saturated''
levels of chromospheric emission, the level is lower than
that for stars hotter than M2.5. The cooler subsample is similar to the
hotter subsample in that there is no strong evidence for
``supersaturation'', but in the cool stars this lack of evidence
extends to $\log N_R \simeq -2.5$ (at least according to our calculation
of the turnover time).

The above results suggest that Rossby number is also not the sole
determining factor for chromospheric activity and that the
limiting level of chromospheric activity at low $N_R$ is spectral type
dependent.  Figure~10 shows the CaT activity index against intrinsic
colour, where we have removed stars with $\log N_R > -1$,
which may be expected to show lower levels of magnetic
activity. Figure~10 shows that the upper  envelope of the
CaT index is clearly colour-dependent. There appears to be a rather
clean break in the data at $(V-I)_0 \simeq 2.3$, corresponding to
spectral types between M2 and M3 (Kenyon \& Hartmann 1995).  Cooler
than this, whilst there is considerable scatter, the mean activity
index falls with increasing colour in a statistically significant way,
reducing by almost a factor of three (0.5\,dex) for the coolest stars
in our sample with spectral types $\sim$M4. Recall that although this
decline could be mitigated by $\sim 0.2$\,dex if the basal photospheric
scaling factor in Fig.~5 were flat, it cannot be removed. 

\begin{figure*}
\centering
\begin{minipage}[t]{0.9\textwidth}
\includegraphics [width = 150mm]{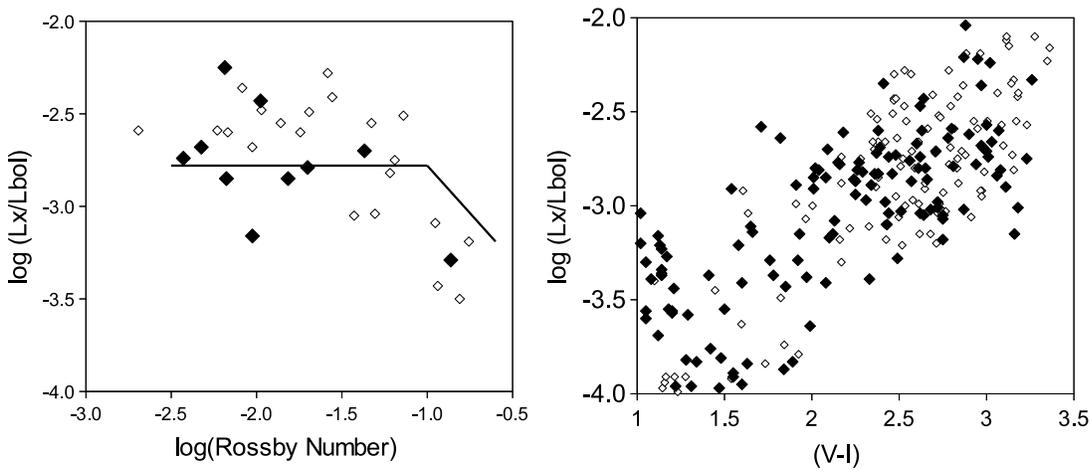}
\end{minipage}
\caption{Variation of $\log(L_{\rm x} /L_{\rm bol})$ with (a) Rossby number and
(b) colour. Values of $\log(L_{\rm x} /L_{\rm bol})$ are derived from X-ray
luminosities for matching targets given in Pillitteri et
al. (2006). Solid diamonds show measured values, open triangles show
upper limits. Plots also show the trend line derived for the CaT (see
Figs. 9 and 10), offset vertically to approximately 
match the saturation levels in $\log(L_{\rm x} /L_{\rm bol})$.}
\label{figxray}
\end{figure*}

\subsection{X-ray activity in NGC~2516}
To assist in deciding whether the fall in the limiting level of
chromospheric emission with decreasing mass represents a reduction in
the efficiency of the magnetic dynamo, or is instead a redistribution of
the activity-related radiative losses to other wavelengths, we examined
the coronal activity of low-mass stars in NGC~2516.

Table 1A of Pillitteri et al. (2006) gives the co-ordinates of
X-ray sources in NGC~2516. These are shown in Fig.~1 along with
our target stars. Unfortunately there is relatively
little overlap. Matching co-ordinates within a radius of 10~arcsec
showed 10 of our NGC~2516 members to be optical counterparts of
identified X-ray sources. A further 21 members were targets for which
Pillitteri et al. give upper limits to their X-ray activity. However,
we have no reason to believe that M-dwarfs towards the centre of
NGC~2516 should have a different rotation period distribution to those
in the outskirts studied by Irwin et al. (2007). It therefore seems
likely that most of the M-dwarfs detected by Pillitteri et al. (2006)
will be in the ''saturated'' regime with low  Rossby numbers.

Figure 11a show $L_{_{\rm x}}/L_{\rm bol}$ as a function of Rossby number
for those stars with known periods. Also represented schematically is
the trend seen in the CaT activity index for K3-M2.5 stars, offset
vertically to match the apparent X-ray saturation level. Two results
lie $\simeq$0.7\,dex above the others possibly due to X-ray
flaring. The remaining results are too sparse to show whether
$\log(L_{\rm x}/L_{\rm bol})$ increases or decreases at very small Rossby
numbers. However there is evidence of saturation for $\log N_R < -1$
and there are two objects with the colours and Rossby numbers of
saturated M4 dwarfs but which do not show any evidence for the
decreased levels of magnetic activity indicated by their CaT indices.
 
Figure 11b shows $L_{_{\rm x}}/L_{\rm bol}$ versus colour for the whole
Pillitteri et al. (2006) sample, including upper limits for photometric
candidates without X-ray detections, some of which may not be cluster members.
The overall picture suggests that the upper envelope of X-ray emission
does not decrease with colour for $V-I>2.5$ as is seen for the CaT
indices (see Fig.~10) and if anything may increase. We must
temper this conclusion by remarking that the number of upper limits
among the data make it impossible to say what the spread in X-ray
activity is at these colours, so that the average X-ray activity may
yet decrease when more sensitive X-ray observations are
available. Nevertheless, that the upper envelope of saturated X-ray
activity does not decrease with decreasing mass, at least as far as
spectral types M5--M6, is reinforced by similar measurements in other open
clusters (Jeffries et al. in preparation) and also from volume limited
samples of field M-dwarfs (Delfosse et al. 1998; Reiners 2008)

\section{Discussion}

In section 5.1 we identified 210 low-mass members of NGC~2516 with
published periods and spectral types between K3 and M4 ($1.1<(V-I)_0 <
3.0$, see Table 2).  These stars have a very high probability of
membership, being identified as probable members by Irwin et al. (2007)
on the basis of their photometry and now also having RVs which are
narrowly distributed arond the mean cluster RV. The estimated intrinsic
velocity dispersion is only $0.66\pm 0.17$~km\,s$^{-1}$ and it seems
likely that we are dealing with a low-mass population that has a
homogeneous age and chemical composition.

The three main results of this study are
\begin{enumerate}
\item Low mass stars with spectral type earlier than M2.5 show
  chromospheric activity, as measured by their CaT activity indices,
  that increases with decreasing Rossby number, reaching a saturated
  plateau for $N_R < 0.1$. The saturated level is roughly independent
  of spectral type between mid-K and M2.5.
\item Almost all the stars in our sample with spectral type cooler 
  than M2.5 rotate fast enough to have $N_R<0.1$. However, the average CaT activity indices
  show a decline towards cooler spectral types of a factor
  of 2--3 between spectral types M2.5 and M4.
\item For stars of all spectral type we see no evidence of chromospheric
  supersaturation. That is, there is no evidence that the CaT activity
  indices fall for the fastest rotators, even for $N_R<0.01$ where
  coronal supersaturation has been claimed for G--K stars.
\end{enumerate}

\subsection{ Saturation of chromospheric activity at low Rossby numbers}

The first of these results is not surprising. Pizzolato et al. (2003)
found that if convective turnover times and Rossby numbers
were calculated according to the simple approximation we have used,
then all stars of spectral types F--K (and a few M-dwarfs) displayed
coronal magnetic activity that behaves in a similar way to the
chromospheric activity we have observed. This includes a saturated
level of activity for $N_R \la 0.1$. The similarity of the threshold
Rossby number for the saturation of coronal and chromospheric emission,
when convective turnover times are calculated consistently, suggests
that the two phenomena are driven by similar physical mechanisms.

\subsection{The dependence of chromospheric activity on spectral type}

The second result is also not new in a qualitative sense.  Mohanty \&
Basri (2003) and West et al. (2004, 2008) have also found that the peak
levels of chromospheric activity in large samples of M-dwarf field
stars, as indicated by $L_{{\rm H}\alpha}/L_{\rm bol}$, tend to fall as
one moves to cooler objects. Where exactly the decline begins is
extremely difficult to ascertain using current field star samples. 
Rapid rotation is rare among field M-dwarfs earlier
than spectral type M3, so most reside in the unsaturated part of the
activity-Rossby number relationship. At later spectral types, field
M-dwarfs are faster rotators on average and many do show ''saturated''
levels of activity which clearly falls by about a factor of three
between spectral types of M4 and M7 (the envelope of $L_{{\rm
H}\alpha}/L_{\rm bol}$ falls from $10^{-3.5}$ to $10^{-4}$, see Fig.~7
in Mohanty \& Basri 2003).  A fall in the limiting level of
chromospheric activity that begun at M2.5 or M3 might be
disguised in field star samples because of the low number stars
with $N_R < 0.1$ at earlier spectral types.

Limiting levels of chromospheric activity can be found in the early-type
M-dwarfs of young open clusters (as we have found here of
course). Comparable measurements of  $L_{{\rm H}\alpha}/L_{\rm
  bol}$ are given for rapidly rotating early M-dwarfs in the
Pleiades (Terndrup et al. 2000). Converting their H$\alpha$ EWs into
$L_{{\rm H}\alpha}/L_{\rm bol}$, we find that these fast rotating
M-dwarfs exhibit a flat upper activity envelope between spectral types M0 and
M3, but as there are only a handful of measurements for stars with
$(V-I)_0>2.6$, we cannot tell if the saturated level declines in
cooler stars.

A further problem with H$\alpha$ measurements is deciding what portion
of the H$\alpha$ flux is chromospheric. The CaT lines are well-behaved
in the sense that chromospheric activity monotonically fills in the
photospheric absorption profile. The behaviour of the H$\alpha$ line is
much more uncertain. The investigations above simply take the
chromospheric component of H$\alpha$ to be the emission flux above a
pseudo-continuum, but it is quite likely that chromospherically
inactive stars exhibit at least some H$\alpha$ absorption (Cram \&
Mullan 1985). For instance, Walkowicz \& Hawley (2009) show that in M3
dwarfs, a modest amount of activity results in deeper H$\alpha$
absorption (with an EW of $\simeq 0.3$\AA), prior to becoming an
emission line at high activity levels. This absorption will become less
significant for cooler stars where the saturated chromospheric
H$\alpha$ lines have much larger emission EWs, but could significantly
increase the deduced chromospheric activity of earlier M-dwarfs where
the most active have emission EWs of $\simeq 2$\AA. A careful account
of this effect may lead to the conclusion that $L_{{\rm
H}\alpha}/L_{\rm bol}$ also begins to decline at M2.5. Alternatively, a
difference in behaviour could be telling us that these features are
formed in two or more chromospheric components in ratios that vary with
spectral type (e.g. Houdebeine 2009).

Uniquely then, the sample in this paper, with its large size, spectral
type range and preponderance of fast rotators, is well placed to to
refine our picture of the behaviour of chromospheric activity in the
most active M-dwarfs. It seems that the peak levels of chromospheric
activity (as measured by our CaT index) do begin to decline at spectral type M2.5.
There is however no evidence that coronal X-ray emission
behaves in the same way. The upper envelope of $L_{\rm x}/L_{\rm bol}$ in
NGC~2516 rises slightly with increasing colour in the same range where
the CaT emission declines. Of course the upper limits
that are still present in the X-ray data for $2<V-I<3$ mean that average
levels of X-ray emission for stars rotating fast enough to be saturated
could yet show a decline. However, this would also require that the
scatter in the X-ray activity levels would have to be several times
larger than the scatter seen in the chromospheric activity. Many of the
M3--M4 objects would need to have activity levels of only $L_{\rm x}/L_{\rm
bol} \sim 10^{-4}$.  Perhaps this might be possible if X-ray activity
among the most rapid rotators shows ``supersaturation'' (see below),
but a recent deep X-ray observation of fast rotating M-dwarfs in
NGC~2547 shows that they do not and that the average level of X-ray
activity in such objects is flat or slightly rising to at least $(V-I)_0
\simeq 2.8$ (Jeffries et al. in preparation).

\subsection{No evidence for chromospheric supersaturation}

Supersaturation is a phenomenon that has only been persuasively
demonstrated in the coronal emission from G- and K-dwarfs (Prosser et
al. 1996; Stauffer et al. 1997). Specifically, James et al. (2000)
searched for evidence of the effect in field and cluster M-dwarfs but
were unable to detect more than one or two examples showing any significant 
reduction in X-ray activity levels
at very low Rossby numbers, partly through a lack of
ultra-fast rotating M-dwarfs. Here we have many M-dwarfs with
$N_R<0.01$, which is comfortably below the threshold that coronal
supersaturation sets in for G- and K-dwarfs. If the chromospheric
activity of M-dwarfs with $N_R<0.01$ was reduced by a factor of two
with respect to M-dwarfs with $0.01<N_R<0.1$, which is the size of the
effect claimed for coronal supersaturation in G- and K-dwarfs, then it
would clearly be seen in our data.

One uncertainty in this work is the lack of theoretical predictions for
turnover times for stars with $M<0.5\,M_{\odot}$. However, even if we
were to cap our turnover times at a maximum value of 70 days, as
predicted for stars of 0.5\,$M_{\odot}$,
this would only increase Rossby numbers for the coolest
stars in our sample by a maximum of 0.3\,dex and many stars would still
have $N_R<0.01$. As an aside, larger Rossby numbers would 
also underpredict the X-ray activity of slowly rotating field
M-dwarfs (Kiraga \& Stepien 2007).

The lack of chromospheric supersaturation in M-dwarfs concurs with a
similar observation for fast-rotating G- and K-type dwarfs in the
IC~2391 and IC~2602 clusters (Marsden et al. 2009). The difference in that work,
is that the fast-rotating G- and K-stars of IC 2391 and IC~2602 
did show evidence for coronal supersaturation.

\subsection{Interpretation and speculation}

A 150\,Myr isochrone from the Siess et al. (2000) evolutionary models
places the transition to a fully convective star at  a mass, $M_{cc}\simeq
0.35$~M$\odot$ and temperatures of 3590K. The models of Baraffe et
al. (2002) give a similar mass for the transition and a temperature of
3480K.  Using the colour/temperature relations of Kenyon \& Hartmann
(1995) this puts the transition at $(V-I)_0=2.30 \pm 0.16$, at about
spectral type M2.5. The decline we have observed in the limiting
levels of chromopsheric emission also begins at spectral type M2.5 and
this also coincides with the increase in rotation rates seen in our own
sample and in samples of field M-dwarfs. 

Whether the fall in chromospheric emission and increase in rotation rates
 are directly linked to the transition to a fully convective core is
 uncertain.
 Firstly, the decline in CaT activity index does not start at
 precise point, rather there is a change from a constant saturated level to a
 declining level between $(V-I)_0 =2.1$ and $2.4$. Secondly, the magnetic
 fields in active stars may reduce $M_{cc}$. Mullan \& MacDonald (2001) showed
 that including an internal magnetic field in evolutionary models could  reduce
 the effective temperature and luminosity of active stars
 and would, if the internal magnetic field were strong enough, reduce
 $M_{cc}$ to as low as 0.1~$M\odot$. In the case of our sample of
M-dwarfs in NGC~2516
  we already know that they have increased radii and reduced effective
temperatures at a given luminosity when compared to standard
(non-magnetic) evolutionary models (see Jackson et al. 2009). This
could be due to the effects proposed by Mullan \& Macdonald (2001) or
perhaps more likely due to extensive coverage by cool magnetic
starspots (eg. Chabrier et al. 2007). Hence, the changes seen in the chromospheric activity (and
the rotation rates in field stars) as a function of spectral type
might be due to an increasing convective fraction rather than a
transition to full convection.

If we proceed on the basis that the decline in the limiting
levels of chromospheric emission and the increase in rotation rates 
occur around the same spectral type as  the transition 
to a fully or near-fully convective core, and that this is  more than a chance
coincidence, we can ask why chromospheric activity declines for
fully convective stars? The evidence from the coronal activity
in NGC~2516 is that it is not necessarily due to a decline in the
efficiency of the dynamo that produces the magnetic flux responsible
for non-radiative heating in the outer atmosphere. This is
supported by measurements of total magnetic flux, which show kilogauss
field covering large fractions of the stellar surface in active field
M-dwarfs at masses both above and below the fully convective transition (Reiners
 2008). The fact that our targets were all detected due to
photometric modulation by magnetic starpsots also shows that strong
magnetic activity of a qualitatively similar nature continues in fully
convective stars.

Instead, it seems more likely that it is the magnetic topology that changes
as the magnetic dynamo shifts from operating predominantly at the
interface between the convection zone and radiative core, to a
distributed dynamo operating throughout a fully convective star
(e.g. Chabrier and K{\"u}ker 2006; Browning 2008). Donati
et al. (2008) and Morin et al. (2008) have shown, using Zeeman doppler
imaging, that M-dwarfs with radiative cores have modest large-scale
fields with predominantly toroidal and non-axisymmetric poloidal
fields, whereas fully convective M-dwarfs have stronger large-scale
fields that are almost fully poloidal and axisymmetric.  Reiners \&
Basri (2009) point out that most of the magnetic energy is still in
small-scale fields that are unresolved by Zeeman doppler imaging for
both groups of stars, but there is a clear shift towards larger-scale
ordered fields in the fully convective stars.

The transition from an interface to a turbulent convective boundary
means that the chromospheric emission may trace altogether different
structures in each regime. One could speculate that the loss of some
magnetic flux on small scales also results in the loss of regions
responsible for bright chromospheric emission, whilst at the same time
having little effect on the coronal volume and density.

\section{Conclusions} 

We have measured rotation rates and RVs, which confirm 210 late-K to
mid-M dwarfs as members of the open cluster NGC~2516.  These were
previously identified as photometric members by Irwin et al. (2007) and
have measured rotation periods. The RVs of cluster members are tightly
bunched about the mean showing an intrinsic dispersion of
$0.65\pm0.17$~km\,s$^{-1}$.  The projected equatorial velocities show an
increase in the proportion of fast rotators for later spectral types,
such that 90 per cent of M4 stars have $v\sin
i > 15$~km\,s$^{-1}$. Fewer stars are rapid rotators at earlier
spectral types, but still a much greater proportion than are found in a
field star sample, which is expected because the cluster is younger
than the probable spindown timescales for all M-dwarfs.

We have gauged chromospheric activity using intermediate resolution
spectroscopy of the first two of the near infrared calcium triplet
lines (8498\AA\ and 8542\AA) and a spectral subtraction technique to
remove the photospheric contribution. The chromospheric activity of
stars that are hotter or cooler than spectral type M2.5 ($(V-I)_0
\simeq 2.3$) show a differing dependence on rotation
period and colour/mass. Our main findings are: (i) Stars with spectral
type earlier than M2.5 behave like other samples of young G- and K-type
stars. Their chromospheric activity increases with decreasing period,
or decreasing Rossby number, and reaches a saturated plateau for Rossby
numbers smaller than about 0.1. (ii) Cooler stars almost all rotate fast
enough that they have Rossby numbers less than 0.1. For these stars we find
almost no dependence on rotation period or Rossby number, but a rather steep decline in
chromospheric activity with colour, amounting to factors of 2--3
between spectral types M2.5 and M4 ($2.3 < (V-I)_0 < 2.9$). (iii) There
is no evidence in our data for any systematic fall in activity at very
low Rossby numbers ($\la 0.01$), a phenomenon that has been seen in the
coronal emission from fast-rotating G- and K-stars and dubbed
``supersaturation''.

It is tempting to identify changes in the properties of
chromospheric activity with the disappearance of the radiative core of
a star at $\simeq 0.35\,M_{\odot}$ and presumably the emergence of a
new, distributed or turbulent dynamo that operates in fully convective
M-dwarfs. We do not see any corresponding decline in peak levels of
X-ray emission across the fully convective boundary in NGC~2516 and
none has been reported in field stars. This, combined with literature
suggesting that magnetic flux continues to be generated strongly in
fully convective stars leads us to favour a changing magnetic topology
as the cause of both the decline in chromospheric emission and the
rapid increase in angular momentum loss timescales as stars approach
or cross the fully convective boundary.

\section*{Acknowledgments}
Based on observations collected at the European Southern
Observatory,Paranal, Chile through observing programs 380.D-0479 and
266.D-5655. RJJ acknowledges receipt of an STFC studentship.

\nocite{Allen1973a}
\nocite{Baraffe2002a}
\nocite{Bagnulo2003a}
\nocite{Bessell1987a}
\nocite{Browning2008a}
\nocite{Browning2010a}
\nocite{Baliunas1995a}
\nocite{Busa2007a}
\nocite{Carpenter2001a}
\nocite{Cenarro2001a}
\nocite{Cenarro2002a}
%%\nocite{Chabrier2005a}
\nocite{Chabrier2006a}
\nocite{Chabrier2007a}
\nocite{Claret1995a}
\nocite{Cram1985a}
\nocite{Cox1955a}
\nocite{Cutri2003a}
\nocite{Damiani2003a}
\nocite{DAntona2000a}
\nocite{Delfosse1998a}
\nocite{Donati2008a}
\nocite{Gilliland1986a}
\nocite{Gonzalez2000a}
\nocite{Hawley1999a}
\nocite{Horne1986a}
\nocite{Houdebine2009a} 
\nocite{Irwin2007a}
\nocite{Jackson2009a}
\nocite{James1997a}
\nocite{James2000a}
\nocite{Jeffries1997a}
\nocite{Jeffries1998a}
\nocite{Jeffries2001a}
\nocite{Jeffries2007a}
\nocite{Jenkins2009a}
\nocite{Kenyon1995a}
\nocite{Kim1996a}
\nocite{Kiraga2007a}
\nocite{LopezMorales2007a}
\nocite{Lyra2006a}
\nocite{Mallik1994a}
\nocite{Mallik1997a}
\nocite{Mangeney1984}
\nocite{Marsden2009a}
\nocite{Mohanty2003a}
\nocite{Montes1998a}
\nocite{Morales2009a}
\nocite{Moraux2005a}
\nocite{Morin2008a}
\nocite{Mullan2001a}
\nocite{Noyes1984a}
\nocite{Pillitteri2006a}
\nocite{Pizzolato2003a}
\nocite{Prosser1996a}
\nocite{Queloz1998a}
\nocite{Randich1996a}
\nocite{Ribas2008a}
\nocite{Rieke1985a}
\nocite{Reiners2008a}
\nocite{Reiners2008b}
\nocite{Reiners2009a}
\nocite{Reiners2009b}
\nocite{Reiners2010a}
\nocite{Siess2000a}
\nocite{Schiavon1997a}
\nocite{Soderblom1993a}
\nocite{Stauffer1994a}
\nocite{Sung2002a}
\nocite{Terndrup2000a}
\nocite{Terndrup2002a}
%%\nocite{Udalski2003a}
\nocite{vanLeeuwen2009a}
\nocite{Ventura1998a}
\nocite{West2004a}
\nocite{West2008a}
\nocite{Walkowicz2009a}

\bibliographystyle{mn2e} 
\bibliography{RJJbib}

%%%%%%%%%%%%%%%%%%%%%%%%%%%%%%%%%%%%

\bsp % ``This paper has been produced using the ...''

\label{lastpage}

\end{document}